# Electrostatic Effects of Self-Trapped Holes in β-Ga$_2$O$_3$ Devices

Nathan Wriedt[1, a)], Joe McGlone[1,2] Davide Orlandini[1], Siddharth Rajan[1,3 a)]

[1]*Department of Electrical and Computer Engineering, Ohio State University, Columbus, Ohio 43210, USA*

[2]*Institute for Materials Research, 201 W 19th Avenue, Columbus, Ohio 43210, USA*

[3]*Department of Material Science and Engineering, Ohio State University, Columbus, Ohio, 43210, USA*

β-Ga$_2$O$_3$ is an ultra-wide bandgap semiconductor with exceptional properties for power electronics and UV-C optoelectronics, but its behavior under illumination remains poorly understood. In this work, we investigate how optically generated self-trapped holes influence electrostatics and current conduction in β-Ga$_2$O$_3$ devices. Using a vertical Schottky photodiode with a semi-transparent Ni anode, we performed capacitance–voltage (C–V), current–voltage (I–V), and temperature-dependent I–V measurements under dark and above-bandgap illumination. Analysis of photocurrent gain reveals that conventional image-force barrier-lowering models require unrealistically high interfacial electric fields, suggesting the presence of an alternative mechanism. By applying Fowler–Nordheim tunneling theory, we reconcile measured photocurrents and photo-capacitance results with physically plausible fields and quantify the two-dimensional concentration of self-trapped holes. Our findings demonstrate that illumination-induced charge significantly alters device electrostatics. Understanding this tunneling-based photocurrent gain mechanism is critical for designing β-Ga$_2$O$_3$ devices for UV-C detectors and power electronics.

β-Ga$_2$O$_3$ is a promising new semiconductor material. It has several attractive attributes— a Baliga Figure of Merit that is higher than that of its competitors SiC and GaN, [1] several shallow n-type dopants with usable solubilities, [2] an ultra-wide bandgap of 4.5—4.8 eV, [3, 4] and a critical breakdown field of 6—8 MV/cm. [5] Despite its many strengths, β-Ga$_2$O$_3$ is not without limitations. Two principal challenges hinder its development: intrinsically low thermal conductivity, and lack of p-type material. The lack of shallow acceptor dopants precludes the formation of conventional p-n junctions, thereby requiring device designers to use unique architectures to perform those functions.

While thermally ionized holes are challenging to achieve in β-Ga$_2$O$_3$, holes are known to form under illumination. When photons are absorbed by β-Ga$_2$O$_3$, electron-hole pairs are generated. [6] The electrons, which have a low effective mass and a mobility up to 200 cm$^2$/Vs, [7] behave similarly to conventional semiconductors, but the holes localize in polaronic states commonly called 'self-trapped holes.' [8] This happens because the electron-phonon coupling is so strong that the lattice is distorted, effectively creating a quantum well in the valence band that the holes occupy. [9] Reports in literature indicate a hole

a)   Authors to whom correspondence should be addressed: wriedt.1@buckeyemail.osu.edu and rajan@ece.osu.edu

capture time on the scale of $0.5 \times 10^{-12}$ s at room temperature,[8] and the effects of these holes can be observed in photocurrent measurements after the illumination is removed.[10] This asymmetry in formation and recombination time results in a net positive charge in the semiconductor that acts like a fixed charge, bending the bands. A schematic illustration of this process is displayed in Figure 1a. In this work, we show how this change in net positive charge due to illumination can have important effects on electrostatics and current conduction in β-Ga$_2$O$_3$.

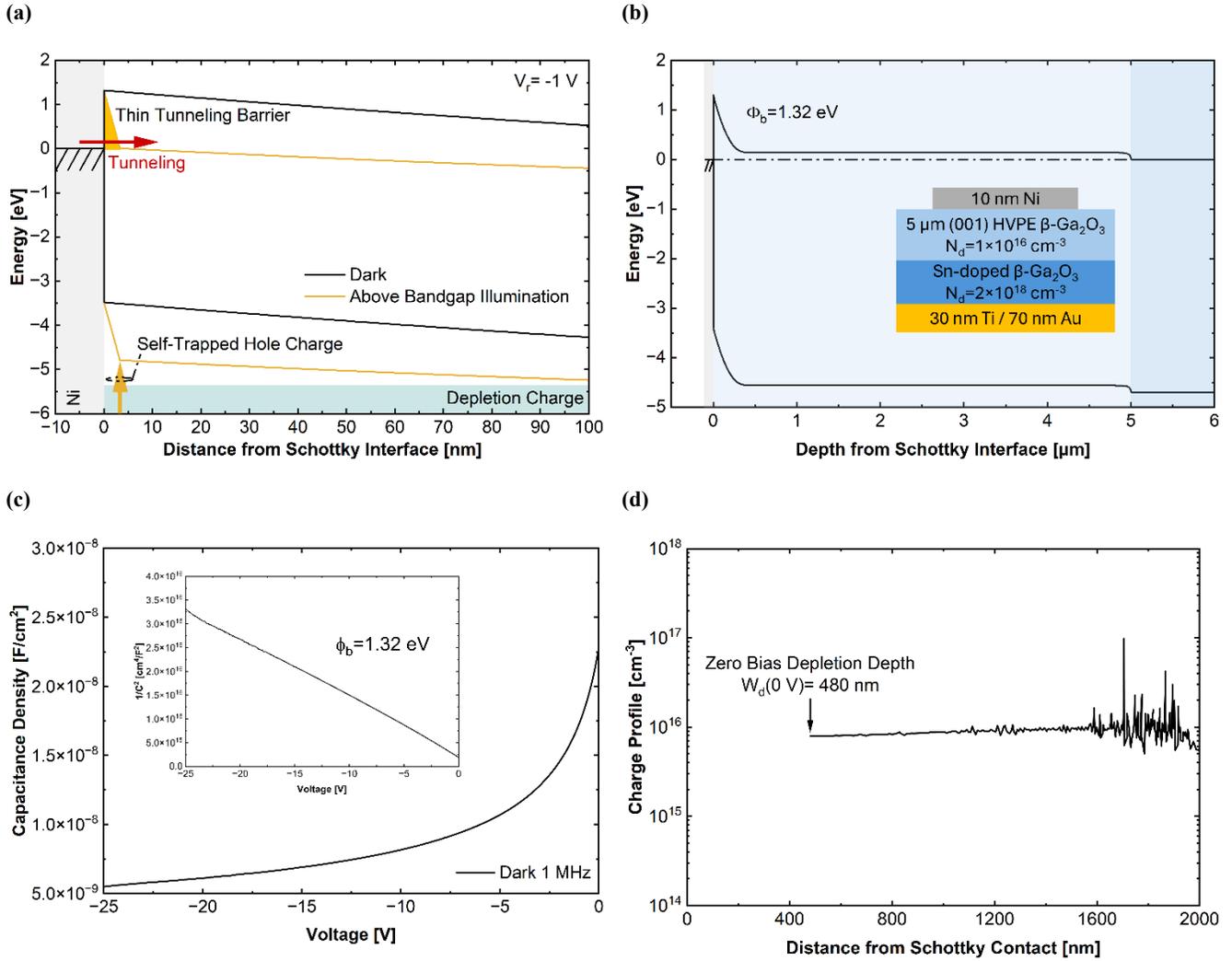

**Figure 1:** (a) A schematic illustration of the mechanism by which self-trapped holes bend the conduction and valence bands in a Schottky diode and (b) equilibrium band diagram of the device under test [inset: schematic diagram], as well as (c) the dark C-V curve [inset: plot of $1/C^2$, showing a built-in voltage of 1.32 eV], and (d) the extracted doping profile ($N_d=1\times10^{16}$ cm$^{-3}$).

To demonstrate the effects of self-trapped holes in β-Ga$_2$O$_3$ optoelectronics, we fabricated a vertical Schottky diode with a semi-transparent anode. We began with an (001)-oriented Sn-doped β-Ga$_2$O$_3$ substrate featuring a 5 μm HVPE (hydride vapor phase epitaxy) epitaxial layer doped at $1 \times 10^{16}$ cm$^{-3}$ from Novel Crystal Technology. Using optical photolithography, we patterned circular regions on the surface with photoresist and deposited 10 nm Ni semi-transparent anodes on these regions via e-beam evaporation. We deposited Ohmic contacts consisting of 30 nm Ti and 70 nm Au on the backside and removed excess Ni through a liftoff process. Figure 1b shows a schematic of the device and presents its equilibrium energy band diagram. To characterize the electrical properties, these devices were subjected to capacitance-voltage measurements under dark conditions; the results are displayed in Figure 1c-d. As shown in the inset of Figure 1c-d, built-in voltage of the device was 1.32 eV and the effective doping density of the HVPE layer was 1X10$^{16}$ cm$^{-3}$.

After characterizing the device in the dark, the devices were subjected to above-bandgap illumination. Using an LED light source, we illuminated the β-Ga$_2$O$_3$ structure with a constant intensity of 110 mW/cm$^2$ at 4.86 eV. We collected photo-capacitance and photocurrent measurements, and the results are shown in Figure 2a and 2b, respectively.

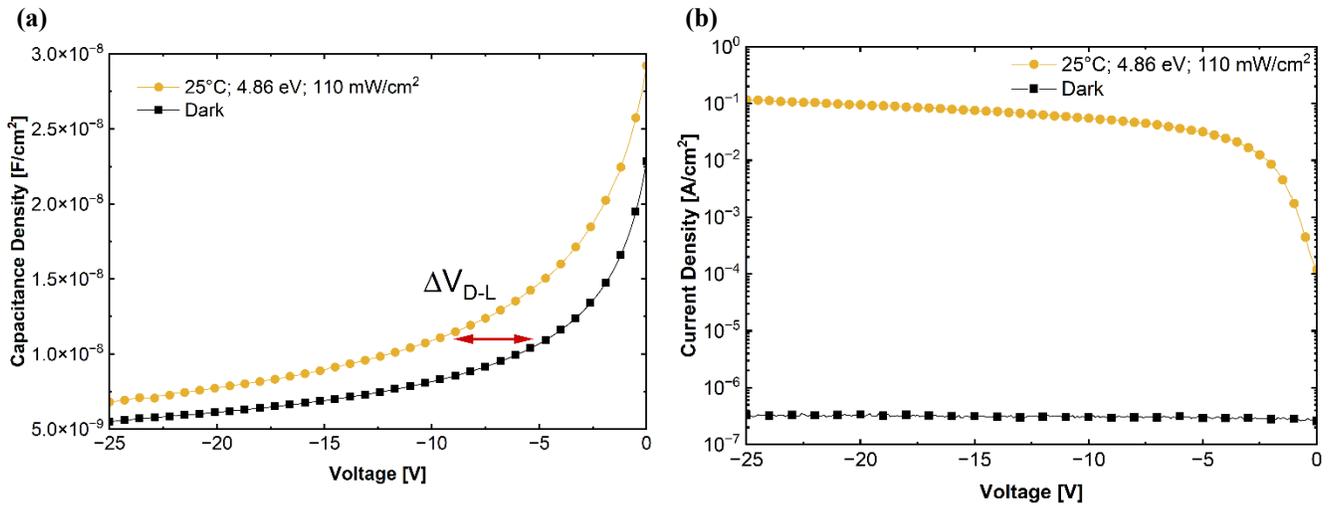

**Figure 2**:(a) Capacitance (1 MHz) and (b) IV measurements collected in both dark and illuminated (4.86 eV; 110 mW/cm$^2$) conditions.

Figure 2a shows the capacitance-voltage curves; under illumination there is a clear leftward shift in the capacitance, which is indicative of excess positive charge.

Based on literature indicating the formation of self-trapped holes under above-bandgap illumination, we believe it is reasonable to assume that the excess charge we see in CV measurements upon illumination with above-bandgap light is caused by an accumulated concentration of self-trapped holes, given their commonly observed and predicted presence in Ga$_2$O$_3$.[8-17] The change in net positive charge causes an effective negative voltage shift in the C-V characteristics:

$$V = \iint_0^{x\prime} [\rho(x)] dx\prime dx \qquad \text{[Equ. 1]}$$

where V is the voltage, x is the depletion length, and $\rho(x)$ is the charge. The effect of a charge distribution can be modeled as the hole sheet density, $\bar{P}_s$ (1/cm$^2$) at a distance centroid, $\bar{x_p}$, from the interface.[6] We can use the difference between the dark and illuminated conditions to find the product of the magnitude of the charge and the distance of the centroid of the charge from the Schottky interface,

$$\Delta V_{D-L} = -\frac{q\bar{P}_s}{\kappa\epsilon_0}\bar{x}_p = -E_{STH}\bar{x}_p \qquad \text{[Equ. 2]}$$

where $\kappa$ is the relative permittivity of the material, $\epsilon_0$ is the permittivity of free space, q is the charge of the electron, and $E_{STH}$ is the electric field due to self-trapped holes. Note that this measurement allows us to estimate the product of the sheet hole density and centroid. To find the location of the centroid or the magnitude of the charge, we need an independent measurement of one of the quantities.

We turn to the LIV curve displayed in Figure 2b, which shows the IV characteristics under 4.86 eV light at an intensity of 110 mW/cm$^2$. This measurement contains much of the data needed to extract the electric field due to self-trapped holes, $E_{STH}$, and the excess trapped hole charge, $\bar{p}_s$, introduced by above-bandgap illumination. However, to extract this information properly, we must know the mechanism by which the self-trapped holes in a vertical Schottky diode produces voltage dependent photocurrent.

To identify the mechanism, we performed temperature-dependent LIV measurements over a range of 25 °C to 150 °C, as shown in Figure 3. The IV curves exhibit a weak, inconsistent temperature dependence: at low voltages, the dependence is negative, but near 13.5 V it becomes positive. The voltage applied to the device is not high enough to produce fields that might induce impact ionization.[18] Instead, such a small and inconsistent temperature dependence suggests that a temperature-independent tunneling mechanism is responsible for the phenomenon.

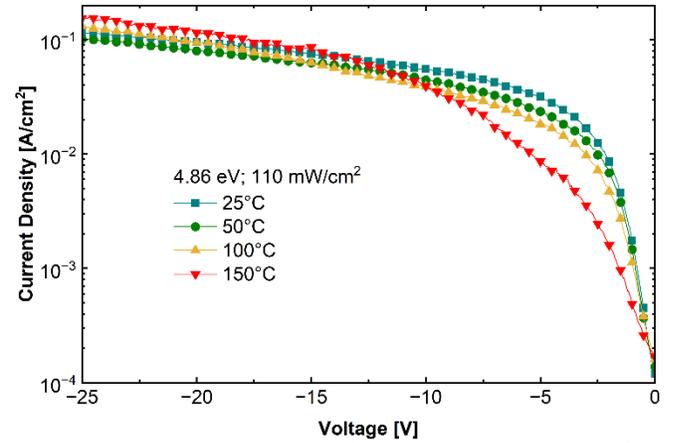

**Figure 3:** Temperature dependent IV curves under 110 mW/cm$^2$ of 4.86 eV CW illumination.

Consider the Fowler-Nordheim equation:

$$J_{FN}(E_{TOT}) = \frac{q^3}{8\pi h\phi_b} E_{max}^2 \exp\left[\frac{-8\pi\sqrt{2m_e}\phi_b^{3/2}}{3hqE_{TOT}}\right] \qquad \text{[Equ. 3]}$$

where $q$ is the charge of an electron, $h$ is the Plank constant, $m_e$ is the effective mass of an electron, $\phi_b$ is the barrier height, and $E_{TOT}$ is the electric field at the interface. Given the barrier height from the dark C-V measurements and the photocurrent from Figure 2b, we can solve Equation 3 for the electric field using the real branch of the Lambert W function:

$$E_{TOT}(J_{FN}(V)) = \frac{8\pi\sqrt{2m_e}\phi_b^{3/2}}{6qhW_0\left[\frac{8\pi\sqrt{2m_e}\phi_b^{3/2}}{6qh}\sqrt{\frac{q^3}{8\pi h J_{FN}\phi_b}}\right]} \qquad \text{[Equ. 4]}$$

Here, $W_0$ is the real branch of the Lambert W function. The results of this analysis, displayed in Figure 4a, represent the electric field on the metal/semiconductor interface. To calculate the magnitude of the excess charge, $\bar{P}_s$, we need to isolate the electric field due to self-trapped holes, $E_{STH}$, by subtracting the electric field present in the device in the dark, $E_{DARK}$, from the total field:

$$E_{TOT} = E_{STH} + E_{DARK} \Rightarrow E_{STH} = \frac{q\bar{P}_s}{k\varepsilon_0} \qquad \text{[Equ. 5]}$$

Figure 4a shows the total field ($E_{TOT}$), dark field ($E_{DARK}$), and field due to self-trapped holes ($E_{STH}$). Using Equation 5, we can find the magnitude of the excess charge, displayed in Figure 4b.

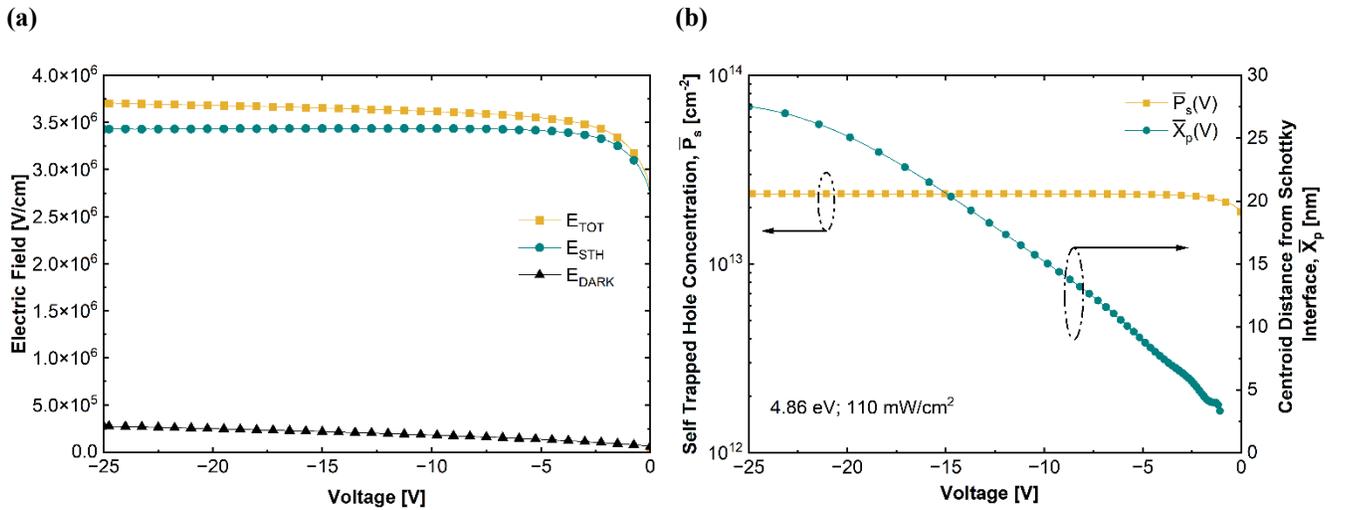

**Figure 4:** Plot of (a) the electric fields observed, extracted from the light-induced current ($E_{TOT}$), due to self-trapped holes, and applied and (b) the magnitude and location of light-induced excess charge with respect to voltage.

Now that we have a value for the magnitude of the excess charge, we can substitute it back into Equation 2, allowing us to solve for the location of the centroid of the charge and the energy band diagram at a given voltage. Figure 4b shows the location of the centroid as a function of applied voltage.

Using the magnitude of the excess charge and the location of the centroid, we can estimate the conduction band profiles. Figure 5a contains the conduction band profile of the device at voltages between -1 and -25 V reverse bias under both light and dark conditions, and Figure 5b contains those same conduction band profiles, but focused on the metal/semiconductor interface.

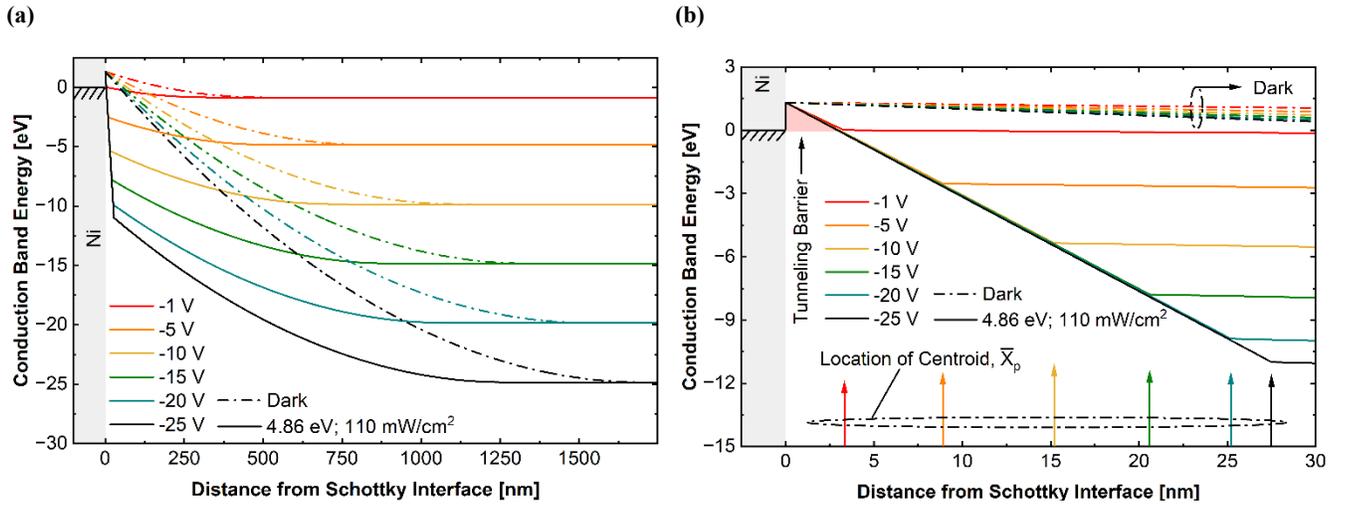

**Figure 5:** Plots of the conduction band profiles at different voltages and under light and dark conditions illustrating (a) the change in depletion region length under illumination caused by the introduction of self-trapped holes assuming that the self-trapped hole distribution is a δ function and (b) the relative lack of change in the size of the tunneling barrier at higher applied reverse biases.

From these two plots, we see that the assumption of F-N tunneling as the dominant conduction mechanism is valid at voltages higher than -1 V reverse bias because the barrier for tunneling is triangular. At voltages higher than -1 V, there is only minor change in the thickness of the tunneling barrier; as the applied voltage increases, the centroid keeps moving further into the device, maintaining the same ~3.25—3.75 MV/cm electric field on the metal/semiconductor interface. Coincidentally, this is also the range in which well-terminated Schottky diodes in β-$Ga_2O_3$ tend to experience a sharp increase in current (in the dark) due to Fowler-Nordheim tunneling. [19-20]

Given this data, we exclude interface states as the source of the observed behavior. We have calculated the charge required to induce the shift in the C-V curve under illumination to be ~$2 \times 10^{13}$ cm$^{-2}$ but also measured a dark current that remains below $3 \times 10^{-7}$ A/cm². Such a high interface-state density which would increase leakage in a Schottky diode above this level.

Additionally, the temperature dependence of the photocurrent does not follow the positive temperature coefficient characteristic of interface traps over the voltage range investigated; instead, the response shows weak, shifting temperature coefficients, consistent with the reduced stability of self-trapped holes at elevated temperatures. Taken alongside literature documenting the self-trapping behavior of holes in Ga$_2$O$_3$, [8-17] these experimental observations support self-trapped holes as the origin of the illumination-induced electrostatic changes.

The magnitude and location of the centroid of the excess charge introduced by above-bandgap illumination can be found using the photo-capacitance and photocurrent measurements, and the experiments suggest that it shifts deeper into the semiconductor with increasing reverse bias at a rate of approximately 1 nm/V (Figure 4b) before slowing down at higher applied voltages. The theory represents the excess charge with a δ-function; however, the actual distribution of holes will be the result of a dynamic equilibrium between electron-hole pair generation, hole transport in the valence band (from semiconductor to metal), and recombination with electrons. Unlike conventional semiconductors, hole transport in β-Ga$_2$O$_3$ displays non-linear dependence on the field due to hole self-trapping with a behavior like hopping transport. At higher applied reverse voltage and field, the rate at which holes are injected into the metal will increase, leading to a reduction in the magnitude of $\bar{P}_s$. A hypothetical schematic illustrating what the reality of this dynamic equilibrium might be is displayed in Figure 6a-c. Further detailed explanation of this dynamic equilibrium will require future study.

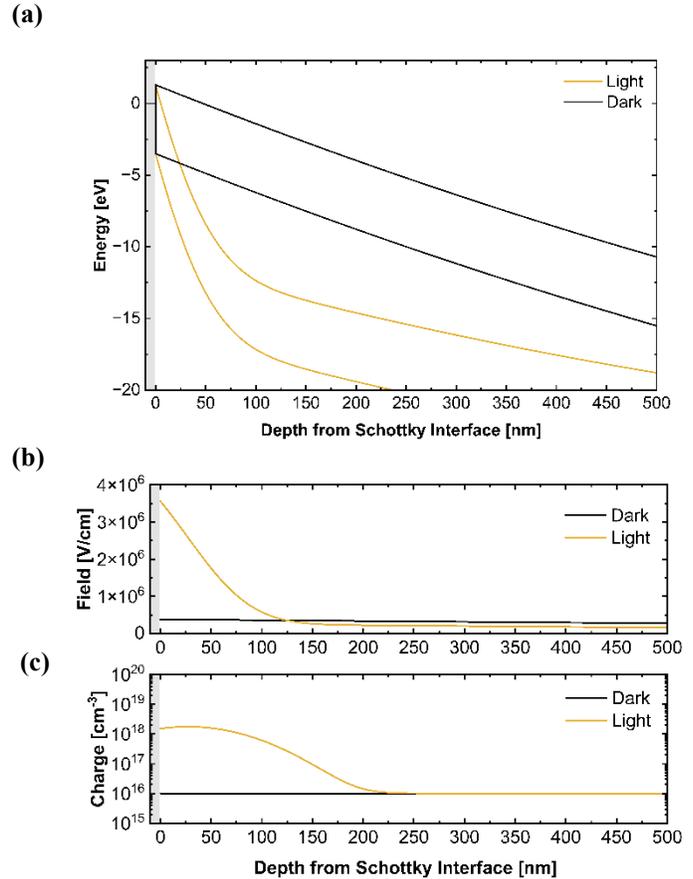

**Figure 6:** Schematic diagram illustrating the (a) energy band diagram and (b) electric field profile resulting from (c) a hypothetical charge distribution that might be formed given the dynamic equilibrium referenced in the text.

To date, the magnitude of photocurrent in β-Ga$_2$O$_3$ vertical Schottky photodetectors is frequently attributed to thermionic emission enhanced by image force barrier lowering caused by self-trapped holes. [21-25] This mechanism is typically described as follows: photons are absorbed by the semiconductor, generating electron-hole pairs. Electrons are collected at the cathode, but holes self-trap, acting similarly to fixed ion charge. [10] This extra charge increases the electric field on the Schottky interface, which then lowers the Schottky barrier. We exclude this argument based on the lack of proper temperature dependence of the photocurrent and the unrealistically high electric fields required to lower the Schottky barrier by the requisite magnitude. According to this mechanism, the photocurrent should follow Equation 6: [6]

$$J_{tot}(V) = \left[e^{\frac{\Delta\phi_b(V)}{k_bT}}\right]J_d^0 + J_{ph}^0 \qquad \text{[Equ. 6]}$$

If this mechanism were driving the photocurrent, we should expect to see a large positive temperature dependence due to the 1/T term in the exponential, but we do not. Furthermore, if this mechanism were the origin of the increased photocurrent, we should use the measured photocurrent from Figure 2b to extract the magnitude of image force barrier lowering according to Equation 7.

$$\Delta\phi_b(V) = k_bT \ln\left[\frac{J_{tot}(V)-J_{ph}^0}{J_d^0}\right] \qquad \text{[Equ. 7]}$$

The magnitude of barrier lowering necessarily requires an electric field on the interface according to Equation 8: [26]

$$E_{TOT}(V) = \frac{4\pi\kappa\varepsilon_0\Delta\phi_b(V)^2}{q} \qquad \text{[Equ. 8]}$$

By applying Equations 6-8, we can create a plot of the interfacial electric field, $E_{TOT}$, versus applied voltage. This analysis was carried out using a relative permittivity of 12.4, [27] and the resulting electric fields required to induce the magnitude of currents measured in Figure 2b are displayed in Figure 7. This interfacial field is higher than the critical electric field of β-$Ga_2O_3$. [5, 18] Given the unrealistically high electric fields required to induce the required magnitude of Schottky barrier lowering and the lack of proper temperature dependence in the photocurrent, we exclude thermionic emission enhanced by image force lowering as a mechanism of the voltage-dependent photocurrent gain in this data.

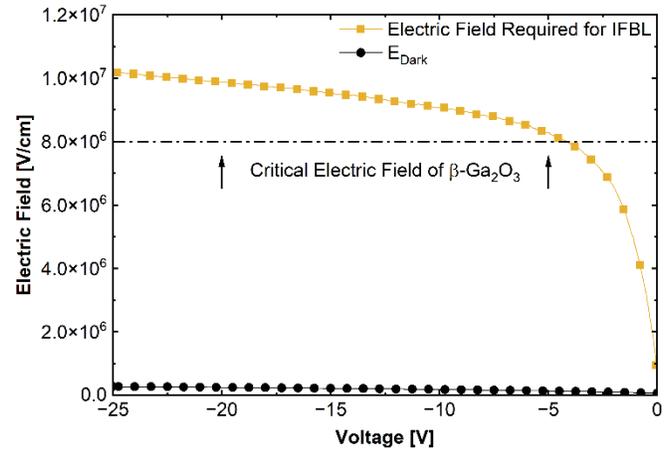

**Figure 7:** Plot of the electric field required to create the photocurrent assuming the mechanism for photocurrent gain present in the device is image force barrier lowering. This plot shows that the electric field required exceeds the critical electric field of $Ga_2O_3$.

In summary, we have shown how charge, field, and energy band profiles of β-$Ga_2O_3$ photodetectors change upon illumination with above-bandgap light because of excess charges introduced by optically generated self-trapped holes based on the evidence provided by I-V, C-V, and I-V-T measurements in dark and illuminated conditions. β-Ga$_2$O$_3$ provides a rather unique case of a semiconductor with excellent transport properties which also has self-trapped holes. While the presence of self-trapped holes in illuminated β-Ga$_2$O$_3$ devices is well-established, [8, 9, 24] this work shows the impact of polarons on device electrostatics and current characteristics. Our results show that the photoconductive gain in β-Ga$_2$O$_3$ Schottky photodetectors originates from electron tunneling from the metal into the semiconductor, rather than from thermionic emission enhanced by image-force barrier lowering. We show that hole-induced tunneling uniquely explains the current–voltage behavior typically reported in Ga$_2$O$_3$ optoelectronic devices, and we provide a quantitative, experimentally validated, physics-based rationale for this mechanism. By adopting a model that includes tunneling of electrons from the metal, we reconcile measured photocurrents with physically plausible electric fields, extract a two-dimensional concentration of self-trapped holes which reshape the internal field profile, and estimate the energy band diagram at a range of voltages. Understanding this tunneling-based gain mechanism and, more fundamentally, the behaviors and impact of self-trapped holes on the electrostatic conditions of β-Ga$_2$O$_3$ devices is essential for understanding the characteristics of electronic and photonic β-Ga$_2$O$_3$ devices. This work is relevant for understanding the behavior of Ga$_2$O$_3$ devices in the presence of holes, whether they are generated by impact ionization or optical generation. Furthermore, the ability to introduce a two-dimensional charge concentration as large as $2\times10^{13}$ cm$^2$ by illuminating the material with above-bandgap light could enable novel device structures. With β-Ga$_2$O$_3$'s other notable advantages as a material for UV-C optoelectronics and power electronics, this phenomenon could be exploited for several new and exciting device applications.

We acknowledge funding from DOE/NNSA under Award Number(s) DE-NA000392, AFOSR GAME MURI (Award No. FA9550-18-1-0479, project manager Dr. Ali Sayir). The content of the information does not necessarily reflect the position or the policy of the federal government, and no official endorsement should be inferred.

The authors have no conflicts to disclose.

The data that support the findings of this study are available from the corresponding author upon reasonable request.